\documentclass[conference,10pt,twocolumn]{IEEEtran} %draftclsnofoot

\usepackage{cite} 
\usepackage{graphicx}
\usepackage[english]{babel}
\usepackage{amsmath}
\usepackage{amsthm}
\usepackage[table,xcdraw]{xcolor}
\usepackage{subfig}
\usepackage[linesnumbered, ruled]{algorithm2e}
\usepackage[noend]{algpseudocode}
\usepackage[font=footnotesize]{caption}
\usepackage{epsfig}
\usepackage{xfrac}
\usepackage{nicefrac}
\usepackage{psfrag}
\usepackage{times}
\usepackage{balance} 
\usepackage{amssymb}
\usepackage{amsfonts}
\usepackage{bm}
\usepackage[bottom]{footmisc}
\usepackage{mwe}
\usepackage{color}
\usepackage{booktabs,multirow}

\SetKwRepeat{Do}{do}{while}
\SetKwInOut{Input}{Input}
\SetKwInOut{Output}{Output}

\begin{document}

%%%%%%%%%%%%%%%%%%%%%%%%%%%%%%%%%%%%%%%%%%%%%%%%%%%%%%%%%%%%%%%%%%%%%%%%%%%%%%%%%%%%%%%%%%%%%%%%%%%%%%
%%%%%%%%%%%%%%%%%%%%%%%%%%%%%%%%%%%%%%%%%%%%%%%%%%%%%%%%%%%%%%%%%%%%%%%%%%%%%%%%%%%%%%%%%%%%%%%%%%%%%%
% \title{Joint Satellite Association and UAV Control for Non-Terrestrial Low-Latency Communication}
% \title{\fontsize{21}{24}\selectfont Joint Satellite and UAV Mobility Management via Machine Learning for Non-Terrestrial Low-Latency Communication}

% \title{Joint LEO Satellite Association and UAV Relaying Control for Non-Terrestrial Network}
% \title{\fontsize{21}{24}\selectfont Joint Satellite and UAV Relaying Mobility Management via Machine Learning for Non-Terrestrial Networks}

\title{\fontsize{21}{24}\selectfont Integrating LEO Satellite and UAV Relaying via Reinforcement Learning for Non-Terrestrial Networks}

% \title{\fontsize{21}{24}\selectfont Dynamic Planning of LEO Satellite and UAV via Machine Learning for Non-Terrestrial Networks}
% \title{\fontsize{21}{24}\selectfont Relaying while Flying: Dynamic Planning of LEO Satellite and UAV via Machine Learning for Non-Terrestrial Networks}
% \title{\fontsize{21}{24}\selectfont Integrating LEO Satellite and UAV Relaying via Machine Learning for Rate-Oriented Non-Terrestrial Communication}

\author{ 
Ju-Hyung Lee$^\dag$, Jihong Park$^*$, Mehdi Bennis$^\ddag$, and Young-Chai Ko$^\dag$ \\

	\small $^\dag$Electrical and Computer Engineering, Korea University,  
	\small Seoul, Korea \\
	\small $^*$ School of Information Technology, Deakin University, Geelong, VIC 3220, Australia\\
	\small $^\ddag$Centre for Wireless Communications, University of Oulu, 90014 Oulu, Finland \\
	\small leejuhyung@korea.ac.kr, jihong.park@deakin.edu.au, mehdi.bennis@oulu.fi, koyc@korea.ac.kr 
	}

\maketitle
\maketitle

%%%%%%%%%%%%%%%%%%%%%%%%%%%%%%%%%%%%%%%%%%%%%%%%%%%%%%%%%%%%%%%%%%%%%%%%%%%%%%%%%%%%%%%%%%%%%%%%%%%%%%
%%%%%%%%%%%%%%%%%%%%%%%%%%%%%%%%%%%%%%%%%%%%%%%%%%%%%%%%%%%%%%%%%%%%%%%%%%%%%%%%%%%%%%%%%%%%%%%%%%%%%%

%%%%%%%%%%%%%%%%%%%%%%%%%%%%%%%%%%%%%%%%%%%%%%%%%%%%%%%%%%%%%%%%%%%%%%%%%%%%%%%%%%%%%%%%%%%%%%%%%%%%%%%%%%%%%%%%%%%%%%%%%%%%%%%%%%%%%%%%%%%%%%%%%%%%%%%%%%%%%%%%%%%%%%%%%%%%%%%%%%%
% ABSTRACT %
%%%%%%%%%%%%%%%%%%%%%%%%%%%%%%%%%%%%%%%%%%%%%%%%%%%%%%%%%%%%%%%%%%%%%%%%%%%%%%%%%%%%%%%%%%%%%%%%%%%%%%%%%%%%%%%%%%%%%%%%%%%%%%%%%%%%%%%%%%%%%%%%%%%%%%%%%%%%%%%%%%%%%%%%%%%%%%%%%%%

\begin{abstract}
A mega-constellation of low-earth orbit (LEO) satellites has the potential to enable long-range communication with low latency. Integrating this with burgeoning unmanned aerial vehicle (UAV) assisted non-terrestrial networks will be a disruptive solution for beyond 5G systems provisioning large-scale three-dimensional connectivity. In this article, we study the problem of forwarding packets between two faraway ground terminals, through an LEO satellite selected from an orbiting constellation and a mobile high-altitude platform (HAP) such as a fixed-wing UAV. To maximize the end-to-end data rate, the satellite association and HAP location should be optimized, which is challenging due to a huge number of orbiting satellites and the resulting time-varying network topology. We tackle this problem using deep reinforcement learning (DRL) with a novel action dimension reduction technique. Simulation results corroborate that our proposed method achieves up to $5.74$x higher average data rate compared to a direct communication baseline without SAT and HAP.
\end{abstract}
%%%%%%%%%%%%%%%%%%%%%%%%%%%%%%%%%%%%%%%%%%%%%%%%%%%%%%%%%%%%%%%%%%%%%%%%%%%%%%%%%%%%%%%%%%%%%%%%%%%%%%%%%%%%%%%%%%%%%%%%%%%%%%%%%%%%%%%%%%%%%%%%%%%%%%%%%%%%%%%%%%%%%%%%%%%%%%%%%%%

%%%%%%%%%%%%%%%%%%%%%%%%%%%%%%%%%%%%%%%%%%%%%%%%%%%%%%%%%%%%%%%%%%%%%%%%%%%%%%%%%%%%%%%%%%%%%%%%%%%%%%%%%%%%%%%%%%%%%%%%%%%%%%%%%%%%%%%%%%%%%%%%%%%%%%%%%%%%%%%%%%%%%%%%%%%%%%%%%%%
% INTRODUCTION %
%%%%%%%%%%%%%%%%%%%%%%%%%%%%%%%%%%%%%%%%%%%%%%%%%%%%%%%%%%%%%%%%%%%%%%%%%%%%%%%%%%%%%%%%%%%%%%%%%%%%%%%%%%%%%%%%%%%%%%%%%%%%%%%%%%%%%%%%%%%%%%%%%%%%%%%%%%%%%%%%%%%%%%%%%%%%%%%%%%%
\section{Introduction} 

We are at the cusp of a revolution where space is envisaged to meet the ground in 5G and beyond \cite{Intro_1, Intro_1_2}. 
Indeed, ground wireless connectivity has already been extending towards the sky, by integrating unmanned aerial vehicles (UAVs) \cite{Intro_2_1, Intro_2_2, Intro_2_3, Intro_2_4}. 
As opposed to fixed ground base stations, these aerial terminals can be mobile and flexibly deployed at traffic hotspots and disaster sites \cite{Intro_3_1, Intro_3_2}, creating three-dimensional (3D) wireless connectivity \cite{Intro_4}.
Witnessing the recent deployment of low-earth orbit (LEO) satellite mega-constellations (e.g., 1,584 Starlink satellites at $550$ km altitude \cite{Starlink}, $2,622$ OneWeb satellites at $1,200$ km \cite{Oneweb}), space is emerging as the new frontier in beyond 5G  communication, wherein LEO satellites, UAVs, and ground terminals are seamlessly integrated, provisioning large-scale 3D wireless connectivity.

\begin{figure}
    % \vspace{-1em}
    \centering
    \includegraphics[width=\linewidth]{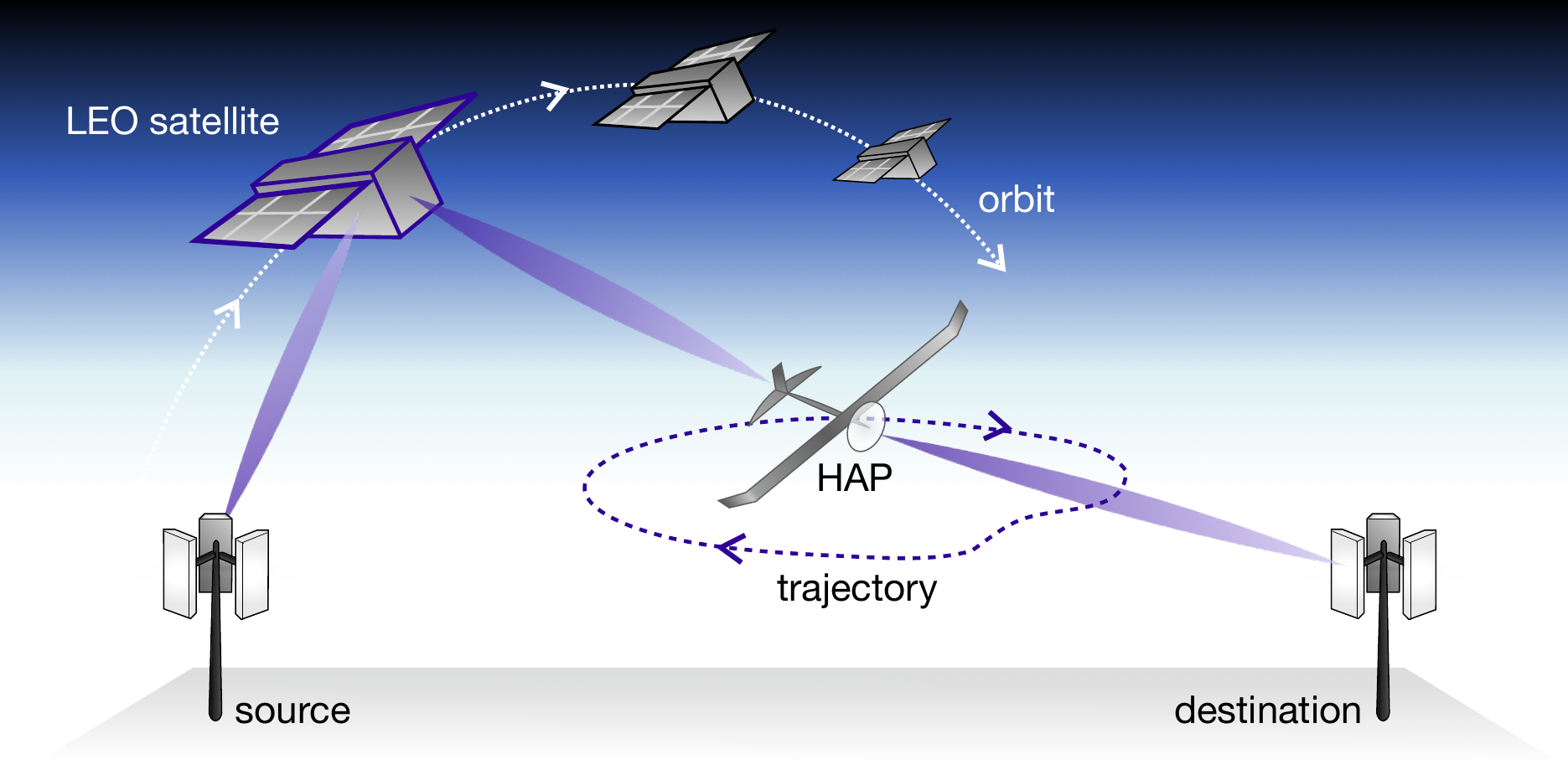}
    \caption{An illustration of source and ground terminals communicating through an orbiting LEO satellite and a moving HAP relay.}
    \label{Illustration}
    % \vspace{-.5em}
\end{figure}

Spurred by these trends, in this article we study the problem of forwarding packets between two faraway ground terminals, \emph{source terminal (Src)} and \emph{destination terminal (Dst)}, relayed through an \emph{LEO satellite (SAT)} and a \emph{high altitude platform (HAP)} such as a fixed-wing UAV or an airship drone, as illustrated in Fig. \ref{Illustration}. 
To maximize the end-to-end (E2E) data rate of the Src-SAT-HAP-Dst link, we aim to optimize the Src-SAT-HAP association while adjusting the HAP location in real time. Solving this problem is non-trivial due to the orbiting SATs and the resultant time-varying network topology.

Concretely, for a fixed network topology, UAV path planning and resource allocation have been studied commonly using the successive convex approximation (SCA) algorithm \cite{SCA_YZ} and dynamic programming under a Markov decision process (MDP) \cite{DP_YZ}. 
These methods become ill-suited for a time-varying topology that is not only difficult to be modelled but also brings about a large and high-dimensional state. 
In addition, due to the mega-constellation of SATs, the number of possible control actions is too huge, incurring high complexity. 

To overcome such challenges, we instead solve the problem using model-free deep reinforcement learning (DRL). 
Furthermore, we reduce the action dimension of Src-SAT-HAP associations, by confining the association candidates only to proximate SATs from Src.

%%% (관련 문제에 대한 기존 해결방식과 이 문제의 challenge) 
% However, the LEO location may change dynamically over time in practice. 
% According to the time-varying LEO locations, the back-haul terminal needs to determine the association and the HAP-relay needs to adjust its trajectory, to ensure the quality-of-service (QoS).
% Recent studies of path planning and resource allocation relies on either dynamic
% programming \cite{} or successive convex approximation (SCA) algorithm \cite{}.
% A major concern lies in that the optimization in \cite{} may not be feasible to deal with the mobile LEO satellite in mega-constellation networks.
% To tackle the issues, Markov decision process (MDP) and reinforcement learning (RL) algorithm can be applied. 
% However, the mobility of LEO satellite and HAP inevitably leads to innumerable states in MDP, making the optimization even more complex. 
% In this context, deep reinforcement learning (DRL) algorithm is more suitable to address the curse of immense state and action spaces induced by the time-varying location of LEO satellite and HAP, than conventional RL algorithm. 

\textbf{Related Works}.\quad 
Towards provisioning high-throughput SAT communication, the industry has recently been deploying mega-constellations of SATs (e.g., SpaceX's Starlink \cite{Starlink}, Amazon's \textit{Kuiper} \cite{Kuiper}, \textit{OneWeb} \cite{Oneweb}, and \textit{Telesat} \cite{Telesat}). Recent works \cite{UCL_1, UCL_2} have advocated that exploiting these SAT relays can achieve faster communication for long distances $>3000$ km than terrestrial optical links. Meanwhile, towards enabling high-throughput non-terrestrial networks (NTNs) \cite{Intro_4}, UAV relay assisted cellular systems have recently been studied, in which the UAV path planning and resource allocation are optimized using successive convex approximation \cite{SCA_YZ, SCA_JH}, block coordinate descent (BCD) method \cite{BCD, BCD_JH}, temporal-difference (TD) method \cite{DP_YZ}. These two trends have been separately investigated, in contrast to our work jointly optimizing SAT and UAV relays. It is noted that in \cite{UCL_2}, the effectiveness of a fixed ground relay in-between SAT links has been studied, as opposed to this work considering a moving UAV relay between SAT links.

\textbf{Contributions and Organization}.\quad 
Our major contributions are summarized as follows.
\begin{itemize}
\item A novel problem has been formulated ($\mathrm{P1}$ in Sec. \ref{Body_1}), which jointly optimizes SAT associations and HAP movement control under the time-varying network topology so as to maximize the average data rate of long-range non-terrestrial communication. To the best of our knowledge, this is the first work taking into account joint satellite and HAP mobility management in the context of non-terrestrial communication.

% We look into the scenario illustrated in Fig. \ref{Illustration}, where multi-hop communications are conducted on LEO satellite with a HAP-aided relaying.
% In order to improve throughput performance in the mega-constellation composed of LEO satellites, an optimal association is determined, and furthermore, a efficient trajectory of HAP (e.g., UAV) is designed according to the time-varying LEO locations.
% To the best of our knowledge, there is no open literature to address the optimization for the high-throughput satellite communication.

\item A DQN based solution has been proposed (see Sec.~\ref{Body}). To cope with a large number of possible actions due to the time-varying network topology, a novel action dimension reduction technique has been applied, which focuses only on a couple of SATs proximal to Src.

% with its action dimension reduction technique

% To tackle the non-convex problems mainly due to time-varying LEO satellite location, we formulate the problem as an MDP model and then introduce proper approach to solve the optimization by adopting a deep $Q$-network (DQN) method to obtain the optimal solution.

\item Numerical results corroborate that our proposed method achieves an average data rate $5.74$x higher than a direct communication baseline without SAT and HAP, $3.99$x higher than the case without HAP. This highlights the importance of not only SAT but also HAP mobility management in enabling high-throughput non-terrestrial communication.

% As a numerical results, the throughput maximized trajectory for HAP-relay is presented.
% \item To reasonably compare the performance of the proposed method, we further find the optimal position of other conventional relaying schemes (e.g., fixed ground-relaying, fixed HAP-relaying), and then we validate the superiority of the proposed compared to baseline and the other relaying schemes.
\end{itemize}

% 이 논문의 구성

The remainder of this article is organized as follows. 
In Sec.~\ref{System Model}, the SAT-HAP assisted non-terrestrial network architecture and channel model are presented. In Sec.~\ref{Body}, the E2E average rate maximization problem is formulated, and a DRL based solution is proposed. In Sec.~\ref{Numerical Result}, simulation results are provided, followed by concluding remarks in Sec.~\ref{conclusion}.

%%%%%%%%%%%%%%%%%%%%%%%%%%%%%%%%%%%%%%%%%%%%%%%%%%%%%%%%%%%%%%%%%%%%%%%%%%%%%%%%%%%%%%%%%%%%%%%%%%%%%%%%%%%%%%%%%%%%%%%%%%%%%%%%%%%%%%%%%%%%%%%%%%%%%%%%%%%%%%%%%%%%%%%%%%%%%%%%%%%

%%%%%%%%%%%%%%%%%%%%%%%%%%%%%%%%%%%%%%%%%%%%%%%%%%%%%%%%%%%%%%%%%%%%%%%%%%%%%%%%%%%%%%%%%%%%%%%%%%%%%%
% SYSTEM MODEL %
%%%%%%%%%%%%%%%%%%%%%%%%%%%%%%%%%%%%%%%%%%%%%%%%%%%%%%%%%%%%%%%%%%%%%%%%%%%%%%%%%%%%%%%%%%%%%%%%%%%%%%
\section{System Model} \label{System Model}

%%% (시나리오)
The network under study consists of an SAT constellation rotating a given orbital plane and a HAP hovering between SAT and Dst, as illustrated in Fig. \ref{Illustration}. 
Specifically, we consider a multi-hop communication link forwarding packets from Src to Dst, through the SAT and HAP relays. 

% In the satellite constellation network, SAT orbits the earth following a certain orbital plane.

%%%%%%%%%%%%%%%%%%%%%%%%%%%%%%%%%%%%%%%%%%%%
\subsection{Dynamics of HAP and SAT}
Based on three dimensional Cartesian coordinates for the location of the terminals, we assume that the back-haul terminal and the terrestrial base station are located at position $\mathbf{q}_{\mathcal{S}}$ and $\mathbf{q}_{\mathcal{D}}$, respectively, while SAT $i$ orbits at the fixed altitude of $H_{\mathrm{L}}$
% \footnote{
% Satellite constellation "Starlink" by \textit{Space X} has launched the vehicle to orbit altitude of $550$ [km] for operation \cite{Starlink}.
% } 
with constant speed $\mathbf{v}_{\mathrm{L}}$ following a predetermined orbital plane.
On the orbital plane, SATs are spaced at equal intervals and circulated.
Specifically, we consider $22$ SATs in the orbital plane and the orbital plane circumstance.
Moreover, we consider that HAP flies horizontally in the $xy$-plane with constant altitude $H_{\mathrm{H}}$.
The time-varying coordinate of the HAP-relay node can be denoted in [km] as  $\mathbf{q}_{\mathrm{H}}(t)=[x(t),y(t),H_{\mathrm{H}   }]^T \in\mathbb{R}^{3 }$, $0\leq t \leq T$.
SATs in one orbit plane and one HAP are considered in this network scenario to reflect a realistic situation.

For ease of analysis, we consider a discrete-time model as in \cite{BCD, BCD_JH}.  
The time horizon $T$ is divided into $N$ time intervals each with duration $\delta_{t}$, i.e., $T=N \cdot \delta_{t}$. 
The duration $\delta_{t}$ is chosen to be sufficiently small so that the HAP's location is adequately approximated within each slot.
Accordingly, the HAP's position $\mathbf{q}_{\mathrm{H}}(t)$ can be approximated in a discrete-time model, i.e., $\mathbf{q}_{\mathrm{H}}[n] \triangleq \mathbf{q}_{\mathrm{H}}(n\delta_{t}) = [x(n\delta_{t}),y(n\delta_{t}),H]^T = [x[n],y[n],H]^T \in\mathbb{R}^{3 }$,  $0 \leq n \leq N+1$.

To obtain a more tractable optimization problem, we apply the discrete linear state-space approximation similarly to \cite{SCA_YZ}. 
Based on the time step size $\delta_{t}$, time $T$ (or $t$) and time slot $N$ (or $n$) can be determined according to $T=\delta_{t} \cdot N$ (or $t=\delta_{t} \cdot n$).
Accordingly, the HAP position $\mathbf{q}_{\mathrm{H}}(t)$ and velocity $\mathbf{v}_{\mathrm{H}}(t)$ can be well characterized by the
discrete-time HAP position vector $\mathbf{q}_{\mathrm{H}}[n] =\mathbf{q}_{\mathrm{H}}(n \delta_{t})$ as well as the velocity vector $\mathbf{v}_{\mathrm{H}}[n] =\mathbf{v}_{\mathrm{H}}(n \delta_{t})$ and the acceleration vector $\mathbf{a}_{\mathrm{H}}[n] =\mathbf{a}_{\mathrm{H}}(n \delta_{t})$ for $n = 0, 1, \cdots, N + 1$.
Thus, the discrete HAP state can be described as
\begin{eqnarray}
\!\!\!\!\! \mathbf{v}_{\mathrm{H}}[n+1] \!\!\!\!&\!=\!&\!\!\!\! \mathbf{v}_{\mathrm{H}}[n]\!+\mathbf{a}_{\mathrm{H}}[n]\delta_t, \ n=1,\cdots, N, \label{C_HAP_v&a} \\
\!\!\!\!\! \mathbf{q}_{\mathrm{H}}[n+1] \!\!\!\!&\!=\!&\!\!\!\! \mathbf{q}_{\mathrm{H}}[n]\!+\mathbf{v}_{\mathrm{H}}[n]\delta_t \!+ \frac{1}{2}\mathbf{a}_{\mathrm{H}}[n]\delta_t^2,  n=1,\cdots, N,  \label{C_HAP_q&v&a}
\end{eqnarray}
where $\mathbf{q}_{\mathrm{H}}[0] = \mathbf{q}_{\mathrm{H}, \mathrm{I}}$ is the initial positions  of the HAP, and $\mathbf{v}_{\mathrm{H}}[0] = \mathbf{v}_{\mathrm{H}, \mathrm{I}}$ is the initial velocity of the HAP.

On the other hand, the discrete state of SAT $i$ in a certain constellation can be expressed as
\begin{eqnarray}
\mathbf{q}^{i}_{\mathrm{L}}[n+1] \!\!&=&\!\! \mathbf{q}^{i}_{\mathrm{L}}[n]+\mathbf{v}_{\mathrm{L}}\delta_t, \ n=1,\cdots, N, \ \forall i,  \label{C_LEO_q&v}
\end{eqnarray}
where the initial position of SAT $i$ in an orbit plane $\mathbf{q}^{i}_{\mathrm{L}}[0] = \mathbf{q}^{i}_{\mathrm{L}, \mathrm{I}}$ and the initial velocity in constellation $\mathbf{v}_{\mathrm{L}}[0] = \mathbf{v}_{\mathrm{L}, \mathrm{I}}$ are determined for each SAT and each orbit plane.

%%%%%%%%%%%%%%%%%%%%%%%%%%%%%%%%%%%%%%%%%%%%
\subsection{Multi-Hop Communication}

We denote a set of SAT in a certain orbital plane as ($\mathcal{I}=\{1,2,\ldots,I\} $) which relays information to a HAP ($\mathcal{K}=\{1\} $).
Two terrestrial terminals, such that one transmits while another receives, are denoted by $\mathcal{S}$ and $\mathcal{D}$, respectively.
In particular, $\mathcal{S}$ refers to a Src and $\mathcal{D}$ refers to a Dst.
RF communication links are used for the aerial-ground channel as well as the satellite-aerial channel
% \footnote{In some industry projects of satellite constellation network \cite{Starlink, Oneweb}, free space optical communication (FSO) is considered to be used in inter satellite/aerial link, which is left for our future work.}
.

%%%%%%%%%%%%%%%%%%%%%%%%%%%%%%%%%%%%%%%%%%%%%%
\subsubsection{Channel Model for RF link}
Since we consider channel characteristics of aerial-ground link or satellite-aerial link, we assume line-of-sight (LoS) links without Doppler effect as in \cite{SCA_YZ}. 
Therefore, the deterministic propagation models are adopted under the position of HAP and attenuation conditions.
The channel gain of RF link $h_{\mathrm{RF}}$ between each terminal at a link distance $d$ can be expressed as $h_{\mathrm{RF}}=\sqrt{\dfrac{\beta_{0}}{d^2}}, \ \forall t,$ where $\beta_{0}$ represents the received power at the reference distance $d_0=1$ [m]. 
Accordingly, the transmission rate in [bps] for the slot $n$ can be expressed as
\begin{equation}
C_{\mathrm{RF}} = B_{\mathrm{RF}} \log_2\left( 1+ \dfrac{\gamma_0}{\| d \|^2}  \right) \ \mathrm{[bps]}, \ \forall t, 
\label{Rate_RF}    
\end{equation}
where $B_{\mathrm{RF}}$ represents the RF bandwidth, and $\gamma_0 = \frac{\beta_{0} \cdot P}{\sigma_{\mathrm{RF}}^2}$ indicates the reference SNR with constant transmission power $P$ and noise variance $\sigma_{RF}^2$.

%%%%%%%%%%%%%%%%%%%%%%%%%%%%%%%%%%%%%%%%%%%%%%
\subsubsection{Link Configuration for Vertical Platform enabled-Multi-Hop Communication}
In this system, a multi-hop network is configured with the help of LEO satellite constellation and HAP.
For instance, information is transmitted from $\mathcal{S}$ to $\mathcal{D}$ via SAT relay node and via HAP relay node, as shown in Fig. \ref{Illustration}.
The instantaneous achievable rates for each link of the multi-hop relay network in time slot $n$ in bps is given by
\begin{eqnarray}
\!\!&\!\!& \mathit{R}_{\mathcal{S}, i}[n] \leq C_{\mathcal{S}, i}[n], \ \forall n,  \label{R_Si}  \\
\!\!&\!\!& \mathit{R}_{i, k}[n] \leq \mathrm{min}\lbrace  C_{i, k}[n], \ \mathit{Q}_{i} + \mathit{R}_{\mathcal{S}, i}[n] \rbrace, \ \forall n, \label{R_ik} \\
\!\!&\!\!& \mathit{R}_{k, \mathcal{D}}[n] \leq \mathrm{min}\lbrace C_{k, \mathcal{D}}[n], \ \mathit{Q}_{k} + \mathit{R}_{i, k}[n] \rbrace, \ \forall n, \label{R_kD}
\end{eqnarray}
where $\mathit{Q}_{i}$ and $\mathit{Q}_{k}$ denote the remaining bits in buffer of $i$ and $k$, respectively. 
Note that the transmission rate of each link in bps are represented, respectively, as 
\begin{eqnarray}
\!\!&\!\!&\!\! C_{\mathcal{S}, i}[n] = w_{\mathcal{S}, i} B_{\mathrm{RF}}
\log_2\left( 1+ \dfrac{\gamma_0}{(d_{\mathcal{S}, i}[n])^{\alpha}}  \right), \ \forall t,  \label{C_Si}  \\
\!\!&\!\!&\!\! C_{i, k}[n] = w_{i, k} B_{\mathrm{RF}}
\log_2\left( 1+ \dfrac{\gamma_0}{(d_{i, k}[n])^{\alpha}}  \right), \ \forall t,   \label{C_ik} \\
\!\!&\!\!&\!\! C_{k, \mathcal{D}}[n] = B_{\mathrm{RF}}
\log_2\left( 1+ \dfrac{\gamma_0}{(d_{k, \mathcal{D}}[n])^{\alpha}}  \right), \ \forall t, \label{C_kD} \end{eqnarray}
in which $w_{\mathcal{S}, i}$ represents the association of $\mathcal{S}$ and SAT constellation in certain orbit plane. 
Note that it holds $w_{\mathcal{S}, i} = w_{i, k}$, since only one SAT is selected per time $t$ (or $n$) for multi-hop communication.
For each RF link, the link distances are expressed as 
\begin{equation}
\begin{split}
& d_{\mathcal{S},i} = \|\mathbf{q}^{i}_{\mathrm{L}}[n] - \mathbf{q}_{\mathcal{S}} \|, \ d_{i,k} = \|\mathbf{q}_{\mathrm{H}}[n] - \mathbf{q}^{i}_{\mathrm{L}}[n] \|,\\
& d_{k,\mathcal{D}} = \|\mathbf{q}_{\mathcal{D}} - \mathbf{q}_{\mathrm{H}}[n] \|, \ \forall i, n. 
\end{split} \label{distance} 
\end{equation}
Note that the position of SAT $\mathbf{q}^{i}_{\mathrm{L}}[n]$ and the position of HAP $\mathbf{q}_{\mathrm{H}}[n]$ are time varying.

%%%%%%%%%%%%%%%%%%%%%%%%%%%%%%%%%%%%%%%%%%%%%%%%%%%%%%%%%%%%%%%%%%%%%%

%%%%%%%%%%%%%%%%%%%%%%%%%%%%%%%%%%%%%%%%%%%%%%%%%%%%%%%%%%%%%%%%%%%%%%%%%%%%%%%%%%%%%%%%%%%%%%%%%%%%%%%%%%%%%%%%%%%%%%%%%%%%%%%%%%%%%%%%%%%%%%%%%%%%%%%%%%%%%%%%%%%%%%%%%%%%%%%%%%
%%%%%% Body1 %%%%%%
%%%%%%%%%%%%%%%%%%%%%%%%%%%%%%%%%%%%%%%%%%%%%%%%%%%%%%%%%%%%%%%%%%%%%%%%%%%%%%%%%%%%%%%%%%%%%%%%%%%%%%%%%%%%%%%%%%%%%%%%%%%%%%%%%%%%%%%%%%%%%%%%%%%%%%%%%%%%%%%%%%%%%%%%%%%%%%%%%%
\section{Optimal Association and Trajectory Design for Throughput Maximization in LEO Satellite Constellation Network} \label{Body}

In this section, we address the optimization of HAP-aided SAT constellation network (e.g., multi-hop relay network) to maximize the end-to-end rate. 
Only one SAT in constellation can forward information between the source and destination as a relay node, while SATs circulate rapidly in the predetermined orbital plane.
Hence for efficient relaying, appropriate node-to-node association decisions are necessary
% \footnote{
% Efficient handover scheme is also necessary, which is out of scope in this paper.
% }
.
Besides, for the HAP which supports the SAT constellation network as another mobile relay node, proper consideration for dynamics of HAP is necessary for efficient relaying.
Therefore, we focus on the optimization of association of $\mathcal{S}-i$ (or $i-k$) and HAP trajectory design.

%%%%%%%%%%%%%%%%%%%%%%%%%%%%%%%%%%%%%%%%%%%%%%
\subsection{Problem Formulation} \label{Body_1}
The following problem, P1, corresponds to a sum throughput maximization under the constraint related to an actual flight condition of HAP.
For mathematical convenience, we define the set of association as $\mathcal{W}=\{ w_{\mathcal{S}, i}[n], i\in \mathcal{I}, \forall n \}$, and the set of HAP acceleration as $\mathcal{A}=\{\mathbf{a}_{\mathrm{H}}[n], \forall n \}$.
\begin{eqnarray}
(\mathrm{P1}) \!&\!	\displaystyle \max_{ \scriptsize \begin{array}{c} \scriptsize \mathcal{W}, \mathcal{A}  \end{array} } \!&\!
	\sum_{n=1}^{N} \mathit{R}_{\mathcal{S}, \mathcal{D}}[n] \nonumber \label{P1} \\ 
\!&\! \textrm{s.t} \!&\! \eqref{C_HAP_v&a}-\eqref{C_LEO_q&v}, \eqref{R_Si}-\eqref{R_kD} \nonumber \\
% \!&\!\!&\! \mathbf{q}^{i}_{\mathrm{L}}[n] = \mathbf{q}^{i}_{L, \mathrm{I}} + (n-1)\mathbf{v}_{\mathrm{H}}_{\mathrm{L}}\delta_t , \ \forall i, n, \label{P1_C_LEO} \\
\!&\!\!&\! \|\mathbf{a}_{\mathrm{H}}[n]\| \leq A_{\max}, \ \forall n, \label{P1_C_Amax} \\
\!&\!\!&\! w_{\mathcal{S}, i}[n] \in \{0,1\}, \ \forall n, i, \label{P1_C_w} \\
\!&\!\!&\! \sum\nolimits_{i \in \mathcal{I}} w_{\mathcal{S}, i}[n] \leq 1, \ \forall n, \label{P1_C_W}
\end{eqnarray}
where $N$ denotes the one orbital cycle of satellite and $A_{\max}$ indicates the maximum acceleration.
Note that the constraint in \eqref{C_LEO_q&v} represents the mobility of SATs.
Considering the maneuverability of HAP, the equality constraints \eqref{C_HAP_v&a} and \eqref{C_HAP_q&v&a} characterize the discrete state-space model for HAP, i.e., position of HAP $\mathbf{q}_{\mathrm{H}}[n]$, the velocity of HAP $\mathbf{v}_{\mathrm{H}}[n]$, as well as the acceleration of HAP $\mathbf{a}_{\mathrm{H}}[n]$. 
In addition, to take practical constraint of the aerial vehicle into account, the acceleration of HAP is constrained with the maximum acceleration in \eqref{P1_C_Amax}.
The buffer constraints of relay \eqref{R_Si}-\eqref{R_kD} represents the information-causality constraints (i.e., the condition that relay node can only forward the information that has been previously received from the source.) to consider the practical condition of multi-hop relay network.
For association, the constraints of \eqref{P1_C_w} and \eqref{P1_C_W} represents that Src (e.g., back-haul terminal) can be linked by at most one SAT.

Without losing generality, assuming decode-and-forward (DF) relaying protocol without buffer, the constraints of \eqref{R_Si}-\eqref{R_kD} evolve to $\mathrm{min}\lbrace C_{\mathrm{RF},\mathcal{S}, i}[n], C_{\mathrm{RF}, i, k}[n], C_{\mathrm{RF}, k, \mathcal{D}}[n] \rbrace$.
Accordingly, the instantaneous achievable rates for the multi-hop relay network in time slot $n$ can then be approximated as
\begin{eqnarray}
\!\!\!\!\!\!\!\!&\!\!\!\!\!\!&\!\!\!\!\!\!\!\! \mathit{R}_{\mathcal{S}, \mathcal{D}}[n] \simeq \mathrm{min}\lbrace C_{\mathrm{RF},\mathcal{S}, i}[n], C_{\mathrm{RF}, i, k}[n], C_{\mathrm{RF}, k, \mathcal{D}}[n] \rbrace, \forall n. \label{Rate_E2E}
\end{eqnarray}

%%%%%%%%%%%%%%%%%%%%%%%%%%%%%%%%%%%%%%%%%%%%%%
\subsection{MDP Modeling for Deep Reinforcement Learning}
%%% Figure 2

In this subsection, the DRL formulation for the optimization of multi-hop communication in SAT network is introduced. 
By \eqref{C_HAP_v&a}-\eqref{C_HAP_q&v&a}, the dynamic states of HAP (e.g., $\mathbf{q}_{\mathrm{H}}[n]$ and $\mathbf{v}_{\mathrm{H}}[n]$) hold Markov characteristics. 
As such, we can formulate the optimization problem as a Markov decision process (MDP).
Based on the MDP property, reinforcement learning (RL), especially, deep-RL is used to address the problem.
As in \cite{MDP}, the policy corresponds to the probability of choosing an action according to the current state. 
The optimal policy $\pi^{*}$ is the policy that contributes to the maximal long-term system reward. 
Our goal is to find $\pi^{*}$ to maximize the average long-term system reward.
A natural mapping of (P1) to an MDP model is as follows:

%%%%%%%%%
\subsubsection{Environment}

\begin{figure}
    % \vspace{-1em}
    \centering
    \includegraphics[width=\linewidth]{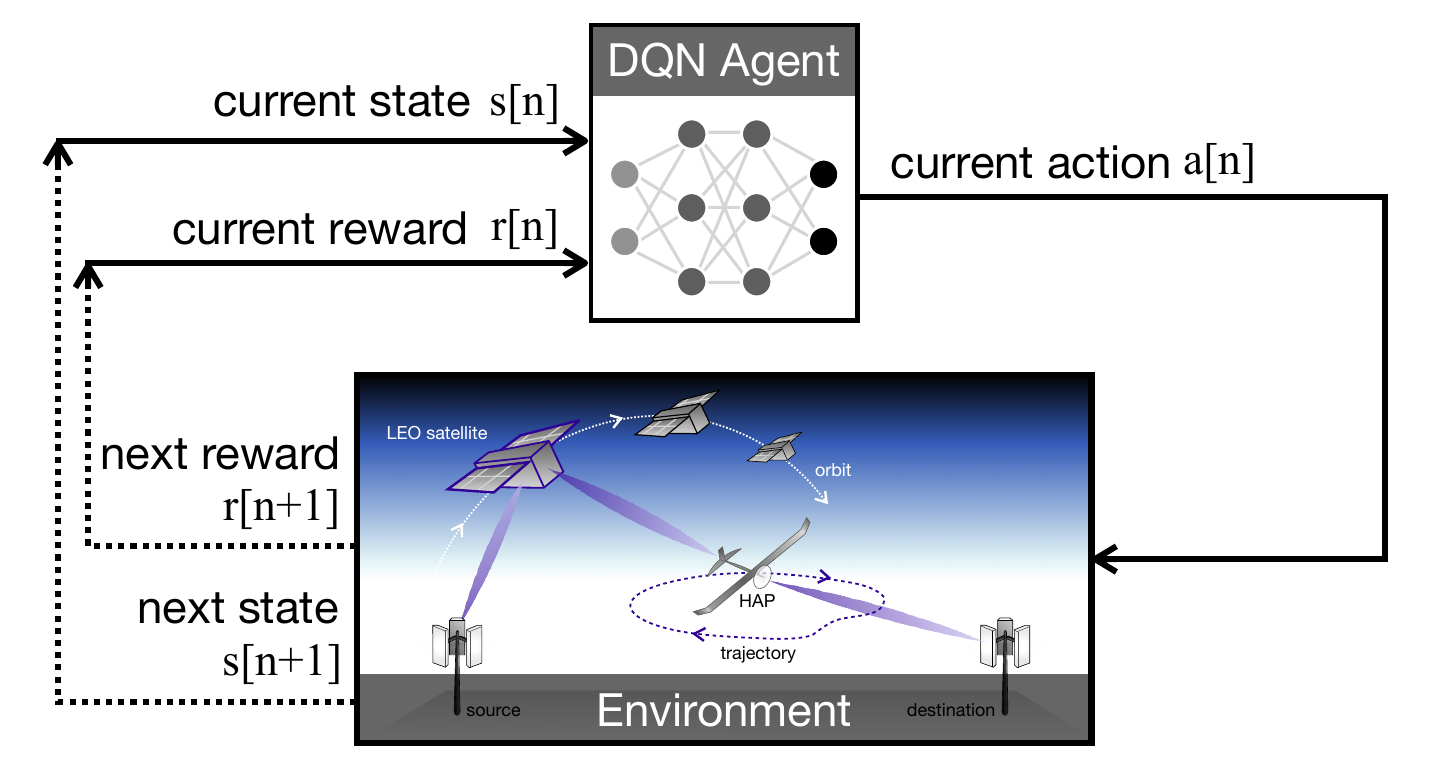}
    \caption{DQN structure for vertical multi-hop communication in LEO satellite constellation network.}
    \label{Structure}
    % \vspace{-1.em}
\end{figure}

%%% Environment 와 observation space 관련
The structure of RL for multi-hop communication in SAT network is shown in Fig. \ref{Structure}, where an agent, corresponding to the HAP and the multi-hop links (e.g., inter satellite/aerial link (ISL)), interact with the environment. 
In this scenario, the environment includes everything outside the ISL link.
At each time $n$, HAP, as the agent, observes a state $s[n]$ from the state space $S$, and accordingly takes an action $a[n]$ from the action space $A$, selecting SAT in the certain orbital plane and acceleration action set based on the policy $\pi$. 
The decision policy $\pi$ is determined by an action-value function $Q(s[n], a[n])$. Following the action, the state of the environment transitions to a new state $s[n+1]$ and the agent receives a reward $r[n]$, determined by the achievable rate in the multi-hop communication.
%%% State space 관련
In our system, the state observed by each node for characterizing the environment consists of several parts: the position of SAT in the orbital plane $\mathbf{q}^{i}_{\mathrm{L}}[n]\in\mathbb{R}^{I \times 3}, i\in \mathcal{I}$, the position of HAP $\mathbf{q}_{\mathrm{H}}[n]\in\mathbb{R}^{3}$, the link distance for each link $d[n] = \lbrace d_{\mathcal{S}, i}[n], d_{i, k}[n], d_{k, \mathcal{D}}[n] \rbrace\in\mathbb{R}^{3}$, the achievable rates for each link including the end-to-end rate (i.e., achievable rate for source-to-destination) $R[n] = \lbrace \mathit{R}_{\mathcal{S}, i}[n], \mathit{R}_{i, k}[n], \mathit{R}_{k, \mathcal{D}}[n], \mathit{R}_{\mathcal{S}, \mathcal{D}}[n] \rbrace \in\mathbb{R}^{4}$. 
Thus the state can be expressed as
\begin{equation}
s[n] = \lbrace \mathbf{q}^{i}_{\mathrm{L}}[n], i\in \mathcal{I}, \mathbf{q}_{\mathrm{H}}[n], d[n], R[n] \rbrace, \ \forall n
\end{equation}

\subsubsection{Action}
Overall, the action space in our system includes two kinds of actions (e.g., $\mathcal{W}$ and $\mathcal{A}$).
Firstly, for the association of $\mathcal{S}-i$, $w_{i, k}[n]$, the source chooses to serve among one of $J$ SAT on the orbital plane in each time slot. 
Equivalently, the decision on $w_{i, k}[n]$ follows $w_{\mathcal{S}, i}[n]$.
Note that our fully observable scenario operates in an offline-manner, so that it is able to optimize $w_{\mathcal{S}, i}[n]$ (and $w_{i, k}[n]$) for each satellite $k$.
To design the association action, $a_{\mathcal{W}}[n] \in\mathbb{R}^{I}$, we use the one-hot encoding.
Secondly, for HAP trajectory design, HAP chooses the acceleration action set in each time slot.
While there are various ways to efficiently handle continuous-state actions, the most straightforward approach is to discretize it to form a finite-state. 
Hence, we consider the acceleration action set, $a_{\mathcal{A}}[n] \in\mathbb{R}^{D \times 2}$, by uniformly discretizing between $0$ and $A_{\max}$.
Note that the acceleration action space can be managed by a discretization level $D$.
Thus, the action space is given as 
\begin{equation}
a[n] = \{ a_{\mathcal{W}}[n], a_{\mathcal{A}}[n] \}, \ \forall n.   
\end{equation}

\subsubsection{Reward}
In the problem, system utility is closely related to the achievable rate in each time slot. 
However, since the correlation between system utility and achievable rate is not linear, we adopt a sigmoid function to describe the correlation as $f(x) = \nicefrac{1}{\left( 1 + e^{-g(x)} \right)}$,
% \begin{equation}
% f(x) = \nicefrac{1}{\left( 1 + e^{-g(x)} \right)},
% \end{equation}
where $g(x) = \frac{x - \mu}{\sigma}$ is the normalization function for $x$.
Note that in the system, $\mu$ is the achievable rate of the baseline (i.e., fixed HAP-relaying in Sec.~\ref{Numerical Result}) and $\sigma$ is the normalization parameter which yields the outcome of $g(R_{\mathcal{S}, \mathcal{D}}[n])$ between $-1$ and $1$.
Thus, via normalization, the reward remains positive if it outperforms a baseline; 
it will be a penalty, a negative reward, if not.
As such, the system reward in the $n$-th time slot induced by the current state $s[n]$ and action $a[n]$ is defined as
\begin{equation}
r[n] = f(R_{\mathcal{S}, \mathcal{D}}[n]), \ \forall n.
\end{equation}
Reward and penalty will drive HAP to find optimal actions which maximize the achievable end-to-end rate.

%%% Episode, Done(terminal state) 관련
With the above MDP model, the objective function of (P1) corresponds to the un-discounted accumulated rewards over an episode up to time slot $N$. 
Note that one episode means that SAT orbits the Earth one time on in its environment.

%%%%%%%%%%%%%%%%%%%%%%%%%%%%%%%%%%%%%%%%%%%%%%
\subsection{Proposed Algorithm with DQN} \label{Body_3}
In the environment in which the state-action space becomes large, many states may be rarely visited by using general $Q$-Learning, thus the corresponding $Q$-values are rarely updated, making convergence difficult.
To overcome the issues, DQN combines $Q$-learning with deep learning \cite{DQN}. 
Basic idea behind DQN is to approximate $Q$-network and update the network weight $\theta$ periodically  (e.g., $Q(s, a; \theta)$). 
A deep neural network provides a mapping between the action-state information and the desired output based on a large amount of training data, which will be used to determine $Q$-values.
DQN updates its weights, $\theta$, at each predetermined iterations to minimize the following loss function with old weight $\theta^{-}$ from the minibatch of replay memory $\mathcal{D}$.
\begin{equation}
L(\theta) = \sum_{(s, a) \in \mathcal{D}} (y - Q(s, a; \theta))^{2},
\end{equation}
where $y = r + max_{a' \in A} Q(s', a'; \theta^{-})$ is the target network.
Note that DQN proceeds towards maximizing the designed reward. 
We follow the detail of DQN algorithm as introduced in \cite{DQN}.

%%%% Action Dimension Reduction technique
% To cope with a large number of possible actions due to the time-varying network topology, a novel action dimension reduction technique has been applied, which focuses only on a couple of SATs proximal to Src.
As identified in \cite{ActionState}, large action spaces can present serious issues for conventional RL algorithms, including DQN.
In our action space, the association action $a_{\mathcal{W}}$ cause the action space to be large, since its dimension follows the number of satellite in the orbital plane.
To clarify the association space dimension related to SAT, we represent the following SATs dynamics in the orbital plane.
Since SATs orbits along a predetermined orbital plane, the satellites in the orbital plane periodically circulate on the surface of Earth.
In other words, these satellites will come back to the same position after a certain amount of time (e.g., orbital period).
Accordingly, the position of SATs, $\mathbf{q}_{\mathrm{L}}[n] \in \mathbb{R}^{I \times 3}$, on the orbital plane is updated as
\begin{equation}
\mathbf{q}_{\mathrm{L}}[n] = \mathbf{q}_{\mathrm{L}}[n] \ \mathrm{mod} \ c_{\mathrm{E}}, \ \forall n,  \label{C_LEO_q&v_1}
\end{equation}
where $\mathrm{mod}$ represents the modulo operation, $c_{\mathrm{E}}=2\pi r_{\mathrm{E}}$ denotes the length of orbital plane (e.g., circumstance of Earth), and $r_{\mathrm{E}}$ denotes the radius of orbital plane. 
Note that each SAT $i$ follows the position as in \eqref{C_LEO_q&v}.

In this regard, we devise the action dimension reduction method for our environment.
It can be seen intuitively that one SAT which located between Src and Dst will be associated, rather than another SAT which is distant from Src or Dst.
Hence, narrowing the view to a specific coordinate space between Src and Dst (e.g., $4000\times4000\times550$ [km$^3$]), a certain pattern is observed that SATs in the coordinate space also periodically circulate.
Considering that SATs in the coordinate space orbit at an equal velocity at equal intervals, we focus on the position of the satellite located proximal to Src in the length of plane in coordinate space $c_{\mathrm{C}}$, instead of the position of all satellites in the whole length of orbital plane.
The position of the selected SATs is thereby rewritten as follows
\begin{equation}
\mathbf{q}_{\mathrm{L}}[n] = \mathbf{q}_{\mathrm{L}}[n] \ \mathrm{mod} \ c_{\mathrm{C}}, \ \forall n, \label{C_LEO_q&v_2}
\end{equation}
Note that $c_{\mathrm{C}}$ depends on the size of considered coordinate space.
As a result, the dimension of association action $a_{\mathcal{W}}[n]$ reduces to $\mathbb{R}^{I'}$ as well as the part of state related to $\mathbf{q}^{i}_{\mathrm{L}}[n] \in \mathbb{R}^{I}$ shrinks to $\mathbf{q}^{i}_{\mathrm{L}}[n], \in \mathbb{R}^{I'}$, since we now only consider the position of selected SAT $\mathbf{q}_{\mathrm{L}}[n] \in \mathbb{R}^{I' \times 3}$.
Note that $I'$ denotes the maximum number of SATs that can be included in a given coordinate space, which depends on $c_{\mathrm{C}}$.
In the following, we present the numerical results of DQN for the optimization of multi-hop communication in LEO satellite network.

%%%%%%%%%%%%%%%%%%%%%%%%%%%%%%%%%%%%%%%%%%%%%%%%%%%%%%%%%%%%%%%%%%%%%%%%%%%%%%%%%%%%%%%%%%%%%%%%%%%%%%%%%
%%%%%%%%%%%%%%%%%%%%%%%%%%%%%%%%%%%%%%%%%%%%%%%%%%%%%%%%%%%%%%%%%%%%%%%%%%%%%%%%%%%%%%%%%%%%%%%%%%%%%%%%%%%%%%%%%%%%%%%%%%%%%%%%%%%%%%%%%%%%%%%%%%%%%%%%%%%%%%%%%%%%%%%%%%%%%%%%%%%

%%%%%%%%%%%%%%%%%%%%%%%%%%%%%%%%%%%%%%%%%%%%%%%%%%%%%%%%%%%%%%%%%%%%%%%%%%%%%%%%%%%%%%%%%%%%%%%%%%%%%%%%%%%%%%%%%%%%%%%%%%%%%%%%%%%%%%%%%%%%%%%%%%%%%%%%%%%%%%%%%%%%%%%%%%%%%%%%%
% Numerical Result %
%%%%%%%%%%%%%%%%%%%%%%%%%%%%%%%%%%%%%%%%%%%%%%%%%%%%%%%%%%%%%%%%%%%%%%%%%%%%%%%%%%%%%%%%%%%%%%%%%%%%%%%%%%%%%%%%%%%%%%%%%%%%%%%%%%%%%%%%%%%%%%%%%%%%%%%%%%%%%%%%%%%%%%%%%%%%%%%%%
\section{Numerical Evaluation} \label{Numerical Result}
In this section, we present the simulation results to demonstrate the performance of the proposed method for throughput maximization in SAT constellation network with mobile HAP-relaying.

In the environment of multi-hop communication, we consider the vertical platforms (i.e., SAT constellation and HAP), in a three-dimensional area of $4000\times4000\times550$ [km$^3$].
For the condition of UAV as a HAP, we assume that the maximum acceleration $A_{\mathrm{max}}=5$ [m/s$^2$].
Unless stated otherwise, we set the time-step size $\delta_t=10$ [s]
The number of neurons per layer is configured as [300,300,200,200] in the deep neural network.
Our DQN model is configured as a five-layer fully connected neural network and three hidden layers. 
The numbers of neurons in the three hidden layers are 300, 300, and 200, respectively. 
The activation function of \textit{tanh} is used.
We follow $\epsilon$-greedy policy to utilize the exploration and exploitation, and use adaptive moment estimation optimizer (Adam) for training. 
The simulation results of this paper are implemented with TensorFlow.
Unless stated otherwise, all parameters are listed in Table \ref{table_Paramter}.

%%%%%%%%%%%%%%%%%%%%%%%% Parameter Table
\begin{table}[t!]\footnotesize
% 	\vspace{.5em}
	\centering
	\caption{Simulation parameters.}
	\resizebox{\columnwidth}{!}{\begin{tabular} {l l}
			%\begin{tabular} {P{1.2cm} P{0.001cm} P{0.001cm} P{0.001cm} P{0.001cm}P{0.001cm} P{0.001cm}}
			\toprule[1pt]
			\textbf{Parameter} & \textbf{Value} \\
			\midrule
			Time slot size & $\delta_t=10$  \\
			Number of time slot for a episode & $N=513$ \\
			\multirow{2}{*}{Positions of Src and Dst} & $\mathbf{q}_{\mathcal{S}}=[0,0,0]^T$, \\  
			& $\mathbf{q}_{\mathcal{D}}=[4000,0,0]^T$ \\
			Altitude of SAT & $H_{\mathrm{L}}=550$ {[}km{]} (from \cite{Starlink}) \\ 
			Velocity of SAT\footnotemark & $\mathbf{v}_{\mathrm{L}, \mathrm{I}}=[0, 7.8, 0]^T$ {[}km/s{]} \\
			Radius of orbital plane & $r_{\mathrm{E}} = 6371$ {[}km{]} \\ 
			Length of orbital plane & $c_{\mathrm{E}} = 40030$ {[}km{]} \\ 
			Number of SAT in orbital plane & $I = 22$ \\ 
			Distance between satellites & $1819$ {[}km{]}  \\
			Length of plane in coordinate space & $c_{\mathrm{C}} = 4000$ {[}km{]} \\ 
			Number of SAT in coordinate space & $I' = 2$ \\ 
			Bandwidth for RF link & $B_{\mathrm{RF}} = 10^{9}$ {[}Hz{]} \\   
			Reference SNR ($d=1$ {[}m{]}), Path-loss exponent & $\gamma_{0} = 10^{9}$, $\alpha = 2$ \\ 
			Learning rate, Discount factor & 0.0001, 0.95 \\ 
			Batch size, \# of iterations/update & 500, 500 \\ 
			Training iterations & 500000 \\
			\bottomrule[1pt]
	\end{tabular}}
	\label{table_Paramter}
% 	\vspace{-1.em}  
\end{table} 

\footnotetext{
To maintain a stable LEO, the orbital velocity is about 7.8 [km/s] at the altitude 550 [km].
}

%%%%%%%%%%%%%%%%%%%%%%%% Figure for Convergence
\begin{figure}
    % \vspace{-.5em}
    \centering
    \includegraphics[width=.88\linewidth]{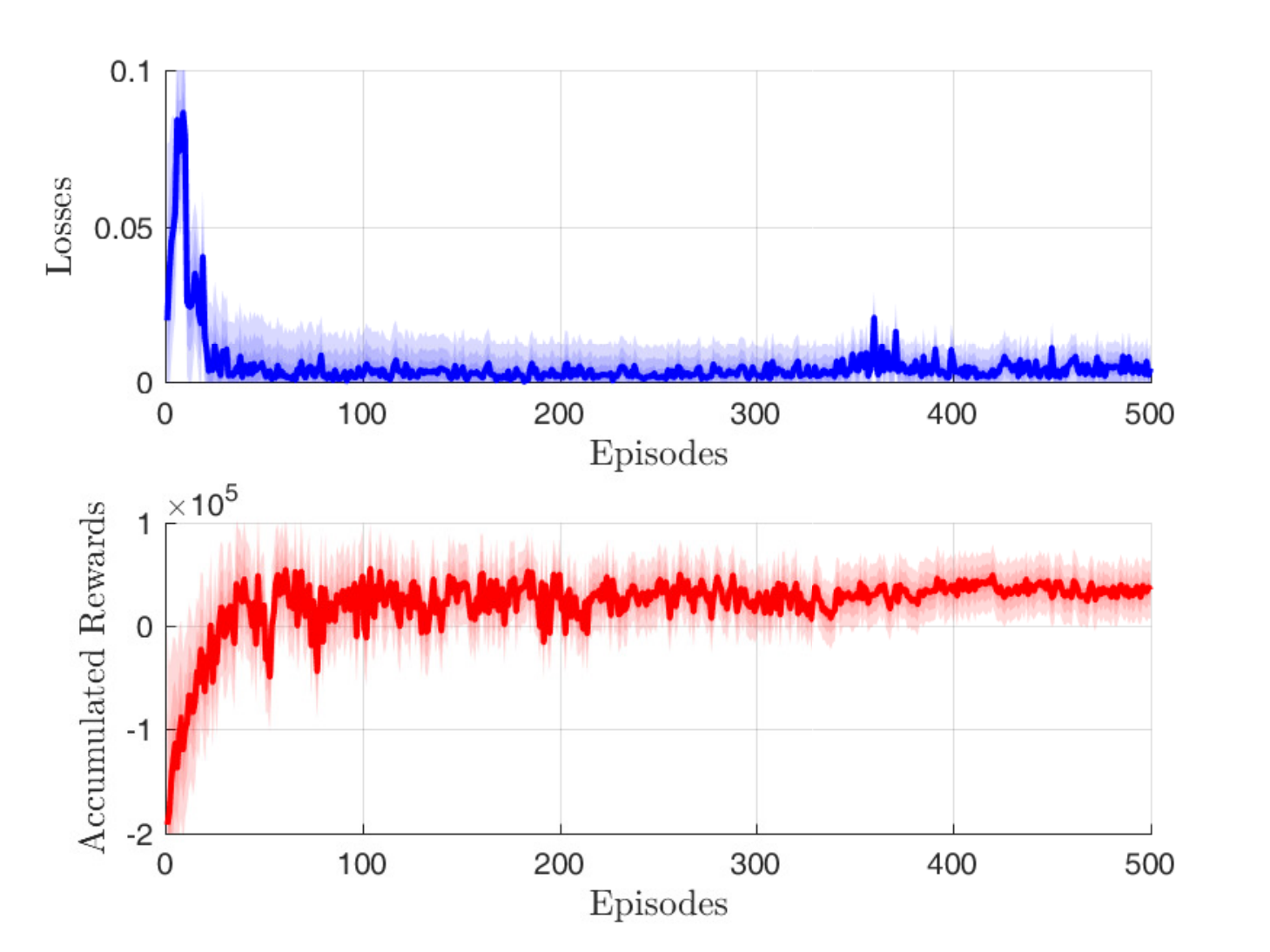}
    \caption{Convergence curve for the proposed algorithm.}
    \label{fig_Convergence}
% 	\vspace{-1.em}
\end{figure}

\textbf{DQN Convergence}.\quad
Fig. \ref{fig_Convergence} depicts the losses and the average reward per episode for the proposed algorithm DQN.
Firstly, for loss curve, it is observed that even with small iteration, the loss value converges to a sufficiently small value.
On the other hand, for the reward curve, it is also observed that the average reward tends to be increasing as iteration goes, while more iterations are required for reasonable convergence.
Note that these curves are the results of averaging four simulation results, and the shaded areas show the fluctuation of the value.

%%%%%%%%%%%%%%%%%%%%%%%% Figure for Optimal Trajectory
\begin{figure*}[t]\centering
\includegraphics[width=.96\linewidth]{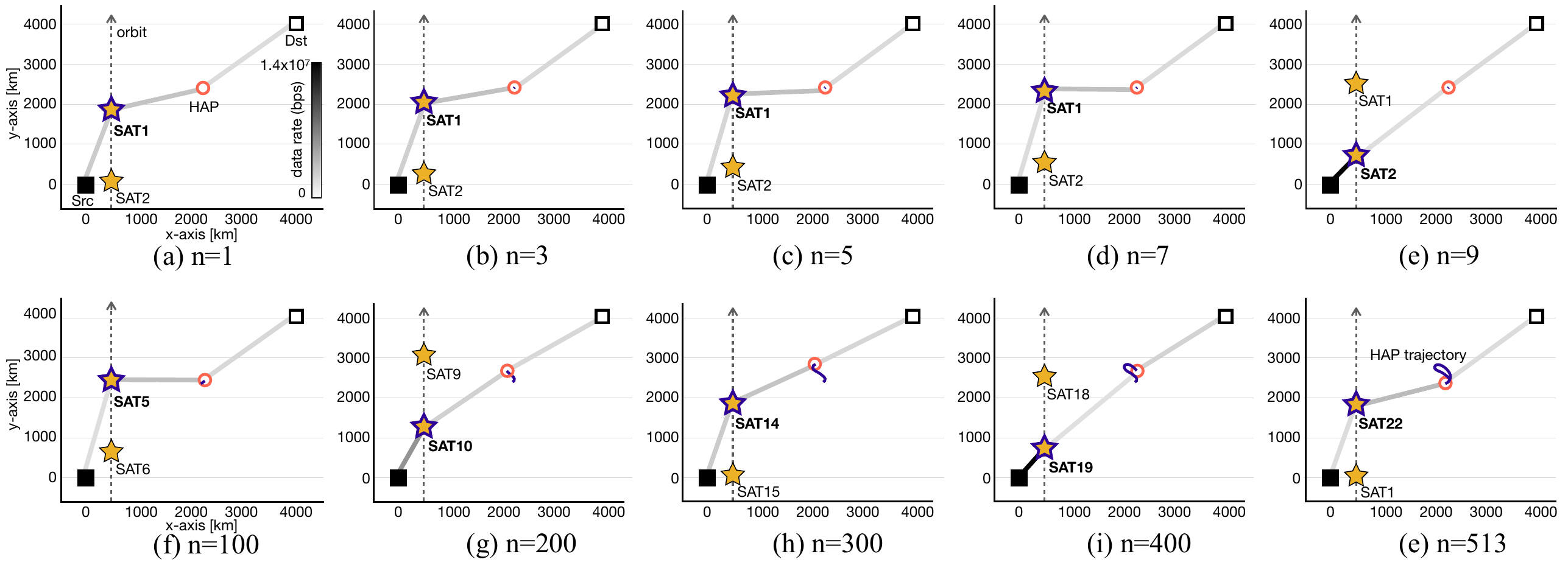}
\caption{Optimized Src-SAT-HAP associations and HAP trajectory during time slots $n=1-513$ (one orbital period).}
\label{fig_Trajectory}
% \vspace{-1.7em}
\end{figure*}

% \begin{figure*}[t]
% % \vspace{-1.5em}
% \begin{minipage}{1\linewidth}
% \centering
% \subfloat[$n = 1$]{\label{fig_n1}\includegraphics[width=.2\textwidth]{Trajectory_N1.eps}}
% \centering
% \subfloat[$n = 8$]{\label{fig_n8}\includegraphics[width=.2\linewidth]{Trajectory_N8.eps}}
% \centering
% \subfloat[$n = 9$]{\label{fig_n9}\includegraphics[width=.2\linewidth]{Trajectory_N9.eps}}
% \centering
% \subfloat[$n = 22$]{\label{fig_n22}\includegraphics[width=.2\linewidth]{Trajectory_N22.eps}}
% \subfloat[$n = 23$]{\label{fig_n23}\includegraphics[width=.2\linewidth]{Trajectory_N23.eps}}
% \end{minipage} \\
% \begin{minipage}{1\linewidth}
% \centering
% \subfloat[$n = 192$]{\label{fig_n192}\includegraphics[width=.2\textwidth]{Trajectory_N192.eps}}
% \centering
% \subfloat[$n = 262$]{\label{fig_n262}\includegraphics[width=.2\linewidth]{Trajectory_N262.eps}}
% \centering
% \subfloat[$n = 360$]{\label{fig_n360}\includegraphics[width=.2\linewidth]{Trajectory_N360.eps}}
% \centering
% \subfloat[$n = 459$]{\label{fig_n459}\includegraphics[width=.2\linewidth]{Trajectory_N459.eps}}
% \subfloat[$n = 513$]{\label{fig_n513}\includegraphics[width=.2\linewidth]{Trajectory_N513.eps}}
% \end{minipage}
% \caption{Optimized Src-SAT-HAP association and HAP trajectory during time slots $n=1-513$ (one orbital period).}
% \label{fig_Trajectory}
% \vspace{-1.7em}
% \end{figure*}

\textbf{Optimal Action for SAT-HAP assisted Non-Terrestrial Network}.\quad
Figs. 4a-4j show the trajectories of the proposed DQN algorithm during one orbital period $n=1-513$.  
Especially in Figs. 4a-4e, the association for Src-SAT-HAP is shown over several time slots.
Although most decision for the association can be intuitively understood, it is worth to note that, Figs. 4c and 4h show that the optimal association differs from an association with the closest SAT from Src. 
Specifically in Fig. 4c, it can be seen that HAP is linked to SAT2, although SAT1 is closer to HAP.
This occurs mainly due to the information-causality constraints $\eqref{R_Si}-\eqref{R_kD}$.
Since E2E data rate of a Src-SAT-HAP-Dst link follows \eqref{Rate_E2E}, the optimal policy decides not to optimize only specific links, but to improve the weakest links.
On the other hand, the designed trajectories for HAP are highlighted in Figs. 4f-4j.
Note that the blue dots in Fig. \ref{fig_Trajectory} represents the trace of HAP.
% We notice that the optimal path was drawn are slightly different each time a simulation is performed, 
The optimal trajectory draws an oval-like shape.
It has a tendency to circulate certain areas between Src and Dst.

%%%%%%%%%%%%%%%%%%%%% Performance Comparison
%% Table 

\begin{table}\footnotesize
% 	\vspace{.5em}
	\centering
	\caption{Spectral efficiencies under different relaying schemes.} 
	\resizebox{\columnwidth}{!}{\begin{tabular} {l c}
			\toprule[1pt]
			\textbf{Relaying Scheme} & \textbf{Spectral Efficiency} [bits/Hz]\\
			\midrule
			Direct Transmission  & \multirow{2}{*}{$ 0.4507 \times 10^{-3}$} \\
			\ without SAT and an Additional Relay & \\
			With SAT & $0.6488 \times 10^{-3}$ \\
			With SAT and a Fixed Ground Relay & $2.2804 \times 10^{-3}$ \\ 
			With SAT and a Fixed HAP Relay & $2.2878 \times 10^{-3}$ \\
			(Proposed) With SAT and Mobile HAP Relay & $\textbf{2.5862} \times 10^{-3}$ \\
			\bottomrule[1pt]
	\end{tabular}}   
% 	\vspace{-1.7em}
	\label{table_Comparison}
\end{table} 

\begin{figure}
    % \vspace{-.3em}
    \centering
    \includegraphics[width=.88\linewidth]{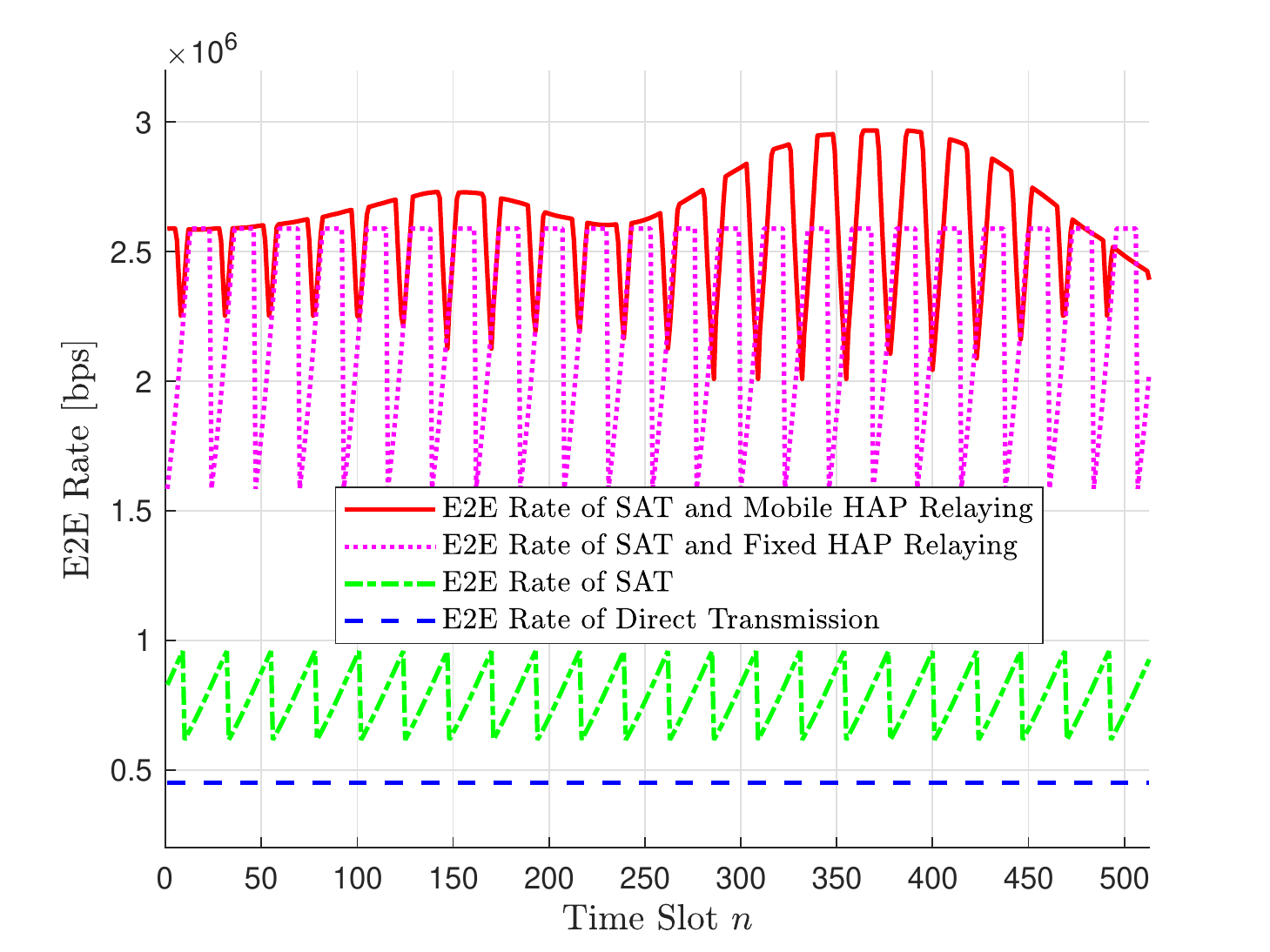}
    \caption{End-to-end data rate for different relaying schemes.}
    \label{fig_E2E}
    % \vspace{-1.7em}
\end{figure}

% \begin{figure}
%     \vspace{-.3em}
%     \centering
%     \includegraphics[width=.88\linewidth]{Fig_Throughput.eps}
%     \caption{
%     Average achievable rates for different relaying schemes.
%     % Performance comparison of relaying schemes for the vertical multi-hop communication in SAT constellation networks.
%     }
%     \label{fig_Comparison}
%     \vspace{-1.7em}
% \end{figure}

\textbf{Impact of Joint SAT-HAP Relaying}.\quad
Our proposed SAT-HAP relaying method is compared with three benchmark schemes;
Firstly, the scenario without any relay node is considered (i.e., Direct transmission without SAT and an additional relaying). 
Secondly, the scenario without any additional relay node only with SAT is considered. 
Lastly, the fixed relaying scheme is considered, in which the fixed relaying with ground terminal (or HAP terminal) is located on the optimal location in the environment.
In the fixed relaying scheme, the optimal position of the fixed relay terminals is found by exhaustive search with a certain grid $95$ [km] as described in the work of \cite{UCL_2}.
Note that we consider the altitude of fixed HAP-relay terminal and fixed ground-relay terminal as $0$ [km] and $H_{\mathrm{H}}=50$ [km], respectively.

Fig. \ref{fig_E2E} compares the E2E rate over time slot, with the three benchmarks.  
We notice that the handover between Src and SAT (i.e. change of association $w_{\mathcal{S}, i}[n]$) cause the E2E rate fluctuation as shown in this figure. 
Considering only one orbit, an edge effect occurs at the end of the episode (e.g., $n = 499-513$), which will be addressed by considering multiple orbits in our future work.
It is found that the proposed mobile HAP-relaying achieves significantly more rate than SAT network without an additional relay terminal.
Furthermore, the proposed solution improves the rate performance reasonably better than the other fixed relaying scheme.

On the other hand, Table. \ref{table_Comparison} compares the spectral efficiency of these schemes.
The proposed mobile HAP-relaying scheme obtain $298.61$ \% rate-gain compared to the scheme without any relay, and also obtain $13.04$ \% rate-gain compared to the fixed HAP-relaying.
Accordingly, the results validate that the additional relay (e.g.,fixed relaying and mobile HAP relaying) can be beneficial for the SAT constellation network in terms of throughput.
Furthermore, these results confirm that using UAV as a HAP-relay terminal has much potential to ensure more throughput via its mobility.

%%%%%%%%%% %%%%%%%%%% %%%%%%%%%%

%%%%%%%%%%%%%%%%%%%%%%%%%%%%%%%%%%%%%%%%%%%%%%%%%%%%%%%%%%%%%%%%%%%%%
%%%%%%%%%%%%%%%%%%%%%%%%%%%%%%%%%%%%%%%%%%%%%%%%%%%%%%%%%%%%%%%%%%%%%
%%%%%% Conclusion %%%%%%
%%%%%%%%%%%%%%%%%%%%%%%%%%%%%%%%%%%%%%%%%%%%%%%%%%%%%%%%%%%%%%%%%%%%%
%%%%%%%%%%%%%%%%%%%%%%%%%%%%%%%%%%%%%%%%%%%%%%%%%%%%%%%%%%%%%%%%%%%%%
\section{Conclusion} \label{conclusion}

%%% 추가적인 relay entity를 위성 LEO 통신 네트워크에 구성하는 것은. 정해진 coverage를 지원하는데, 더 적은 LEO 위성을 deploy 하게끔하여 경제성 측면에서도, reliability와 통신량에서도 도움이 될 거라 기대된다.

In this article, we tackled the problem of maximizing the E2E data rate of a Src-SAT-HAP-Dst link, by jointly optimizing the Src-SAT-HAP association and the HAP location via DQN. Generalizing this single link case to a multi-link scenario could be an interesting direction for future research. To cope with the resulting higher complexity while minimizing additional communication overhead, extending the current DQN based method to a distributed DRL architecture such as actor-critic or multi-agent reinforcement learning (MARL) is an important research topic.

% \vspace{-.5em}

% Configuring additional relay entities in the satellite LEO communication network. It is expected that it will help in terms of economics, reliability, and traffic by supporting fewer coverage and deploying fewer LEO satellites.

%%%%%%%%%%%%%%%%%%%%%%%%%%%%%%%%%%%%%%%%%%%%%%%%%%%%%%%%%%%%%%%%%%%%%%%%%%%%%%%%%%%%%%%%%%%%%%%%%%%%%%%%%%%%%%%%%%%%%%%%%%%%%%%%%%%%%%%%%%%%%%%%%%%%%%%%%%%%%%%%%%%%%%%%%%%%%%%%%%

%%%%%%%%%%%%%%%%%%%%%%%%%%%%%%%%%%%%%%%%%%%%%%%%%%%%%%%%%%%%%%%%%%%%%%%%%%%%%%%%%%%%%%%%%%%%%%%%%%%%%%%%%%%%%%%%%%%%%%%%%%%%%%%%%%%%%%%%%%%%%%%%%%%%%%%%%%%%%%%%%%%%%%%%%%%%%%%%%%
%%%%%% ACK %%%%%%
%%%%%%%%%%%%%%%%%%%%%%%%%%%%%%%%%%%%%%%%%%%%%%%%%%%%%%%%%%%%%%%%%%%%%%%%%%%%%%%%%%%%%%%%%%%%%%%%%%%%%%%%%%%%%%%%%%%%%%%%%%%%%%%%%%%%%%%%%%%%%%%%%%%%%%%%%%%%%%%%%%%%%%%%%%%%%%%%%%
% \section*{ACKNOWLEDGEMENT} 

%%%%%%%%%%%%%%%%%%%%%%%%%%%%%%%%%%%%%%%%%%%%%%%%%%%%%%%%%%%%%%%%%%%%%%%%%%%%%%%%%%%%%%%%%%%%%%%%%%%%%%%%%%%%%%%%%%%%%%%%%%%%%%%%%%%%%%%%%%%%%%%%%%%%%%%%%%%%%%%%%%%%%%%%%%%%%%%%%% BIBLIOGRAPHY %
%%%%%%%%%%%%%%%%%%%%%%%%%%%%%%%%%%%%%%%%%%%%%%%%%%%%%%%%%%%%%%%%%%%%%%%%%%%%%%%%%%%%%%%%%%%%%%%%%%%%%%%%%%%%%%%%%%%%%%%%%%%%%%%%%%%%%%%%%%%%%%%%%%%%%%%%%%%%%%%%%%

\bibliographystyle{IEEEtran} 
% \bibliography{Reference_Paper4_Conf}
\bibliography{Paper4}
%%%%%%%%%%%%%%%%%%%%%%%%%%%%%%%%%%%%%%%%%%%%%%%%%%%%%%%%%%%%%%%%%%%%%%%%%%%%%%%%%%%%%%%%%%%%%%%%%%%%%%%%%%%%%%%%%%%%%%%%%%%%%%%%%%%%%%%%%%%%%%%%%%%%%%%%%%%%%%%%%%%%%%%%%%%%%%%%%

%%%%%%%%%%%%%%%%%%%%%%%%%%%%%%%%%%%%%%%%%%%%%%%%%%%%%%%%%%%%%%%%%%%%%%%%%%%%%%%%%%%%%%%%%%%%%%%%%%%%%%
%%%%%%%%%%%%%%%%%%%%%%%%%%%%%%%%%%%%%%%%%%%%%%%%%%%%%%%%%%%%%%%%%%%%%%%%%%%%%%%%%%%%%%%%%%%%%%%%%%%%%%

\end{document}